\documentclass{PoS}
\usepackage{epstopdf}

\title{A statistical study of fast magnetic reconnection in turbulent accretion disks and jets}

\ShortTitle{Fast magnetic reconnection in turbulent accretion disks and jets}

\author{\speaker{Lu\'{i}s H.S. Kadowaki}\thanks{FAPESP Fellowship.}\\
        Universidade de S\~{a}o Paulo, Instituto de Astronomia, Geof\'{i}sica e Ci\^{e}ncias Atmosf\'{e}ricas \\
	    1226 Mat\~{a}o Street \\
	    S\~{a}o Paulo, 05508-090, Brasil \\
        E-mail: \email{luis.kadowaki@iag.usp.br}}

\author{Elisabete M. de Gouveia Dal Pino\\
        Universidade de S\~{a}o Paulo, Instituto de Astronomia, Geof\'{i}sica e Ci\^{e}ncias Atmosf\'{e}ricas}

\author{Tania E. Medina-Torrej\'{o}n\\
        Universidade de S\~{a}o Paulo, Instituto de Astronomia, Geof\'{i}sica e Ci\^{e}ncias Atmosf\'{e}ricas}
        
\abstract{Fast magnetic reconnection events can play an important role in accretion disk systems. A potential model to explain the non-thermal very-high-energy (VHE) emission (from GeV to TeV) observed in black-hole binaries (BHBs) and Active Galatic Nuclei (AGNs) can be attributed to fast magnetic reconnection induced in the turbulent corona of accretion disks and/or jets. In this work, we will discuss the results of global general relativistic MHD (GRMHD) simulations of accretion disks around black holes, whose turbulence is naturally driven by MHD instabilities, such as the magnetorotational instability (MRI). We will also present studies of magnetic reconnection driven by kink instabilities inside jets employing special relativistic MHD (SRMHD) simulations. As we expect, our simulations reveal the development of a nearly steady-state turbulence driven by these instabilities. We have performed a detailed statistical analysis to identify the presence of current sheets in the turbulent regions of both the accretion flow and jet. We then determined the magnetic reconnection rates in these locations obtaining average reconnection velocities in Alfv\'{e}n speed units of the order of $0.01-0.7$, which are consistent with the predictions of the theory of turbulence-induced fast reconnection.}

\FullConference{International Conference on Black Holes as Cosmic Batteries: UHECRs and Multimessenger Astronomy - BHCB2018\\
	12-15 September, 2018\\
	Foz du Iguazu, Brazil}

\begin{document}
	
\section{Introduction}
	
Accretion disk and jet systems are commonly associated with Black Hole Binaries (BHBs), Active Galactic Nuclei (AGNs) and Young Stellar Objects (YSOs)(see, e.g., \cite{pringle_81, balbus_hawley_98, abramowicz_fragile_13}). High-energy (HE) and very-high-energy (VHE) emissions are frequently observed in BHBs and AGNs. For instance, the X-ray transitions observed in BHBs (see, e.g., \cite{fender_etal_04, belloni_etal_05, remillard_mcClintock_06, kylafis_belloni_15}) are characterized by a high/soft state attributed to the thermal emission of a geometrically thin, optically thick accretion disk \cite{shakura_sunyaev_73} and a low/hard state attributed to inverse Compton of soft X-ray photons by relativistic particles in a geometrically thick, optically thin accretion flow (see \cite{esin_etal_98, esin_etal_01, narayan_mcclintock_08}). Besides, a fast transient state (of the order of a few days; see \cite{remillard_mcClintock_06}) is identified between these two states. VHE emission (gamma-rays in GeV and TeV band) has also been observed in BHBs, such as Cgy-X1 \cite{albert_etal_07} and Cgy-X3 \cite{aleksic_etal_10}. In particular, the origin of the latter is uncertain due to the poor resolution of current gamma-ray detectors. Kadowaki et al. \cite{kadowaki_etal_15} and Singh et al. \cite{singh_etal_15} found that turbulent fast magnetic reconnection \cite{lazarian_vishiniac_99} operating at the coronal region of accretion disks can explain the gamma-ray emission as coming from the core region of BHBs. According to this model, reconnection events between the magnetic field lines lifting from the accretion disk corona and those anchored into the horizon of the black hole could accelerate relativistic particles in a first-order Fermi process (see \cite{dgdp_lazarian_05, dgdp_etal_10, kowal_etal_12, delvalle_etal_16}; see also de Gouveia Dal Pino et al. 2019, in these Proceedings). These particles, interacting with the density, magnetic and radiation fields are able to produce gamma-ray emission. Recently, Khiali et al. \cite{khiali_etal_15} considered the power released by turbulent fast magnetic reconnection events to develop an analytical, single zone scenario to produce the leptonic and hadronic emission to obtain the SEDs of Cgy-X1 and Cgy-X3. The comparison with the observed SEDs shows that this core model assuming magnetic reconnection as a source of acceleration of the particles explains very well the VHE emission of the BHBs (see also \cite{rodriguezramires_etal_18}). 
	
Besides the core region, jets can also be an interesting site for fast magnetic reconnection events. The VHE emission observed from blazars is frequently associated with shocks along the relativistic jets. However, TeV flares have been observed with time-scales of the order of a hundred seconds (see, e.g., MKR 501 and PKS2155-304 \cite{aharonian_etal_07, albert_etal_07b}), pointing to compact emission regions of few Schwarzschild radii.
%which implies in compact emission regions of few Schwarzschild radii.
To avoid the absorption of the VHE emission by pair creation, the emission zone should have a Lorentz factor higher than $50$ (see, e.g., \cite{begelman_etal_08}), but this is much larger than the bulk Lorentz factor of the jets measured in a sub-parsec scale of MKR 501 and PKS2155-304 \cite{giroletti_etal_04, piner_edwards_04, giannios_etal_09}. An acceleration model able to explain such emission was proposed by Giannios et al. \cite{giannios_etal_09}. These authors proposed that relativistic plasmons produced by reconnection (\textit{jets inside jets}) are able to produce the TeV emission and avoid the absorption by pair production with Lorentz factors compatible with those obtained by observations. A reconnection model has also been proposed by Zhang \& Yan \cite{zhang_yan_11} along gamma-ray burst (GRB) jets. In the latter case, fast magnetic reconnection events can be triggered by current driven kink instabilities along Poynting-flux relativistic jets (see \cite{singh_etal_16,duran_etal_17}). This instability excites large-scale helical motions that can strongly distort or even disrupt the system (see \cite{mizuno_etal_09, mizuno_etal_12, mizuno_etal_14, singh_etal_16}).
	
The aim of this work is to probe the viability of turbulent fast magnetic reconnection events in the core regions and jets of compact sources by means of numerical simulations. With this aim, we have performed global general relativistic MHD (GRMHD) simulations of accretion disks around black holes and special relativistic MHD (SRMHD) simulations of Poynting-flux jets and evaluated the presence of turbulent fast reconnection driven by MHD instabilities, such as the magnetorotational instability (MRI; \cite{chandrasekhar_60, balbus_hawley_98, hawley_etal_95}) and current driven kink instabilities \cite{mizuno_etal_09, mizuno_etal_12, mizuno_etal_14}. To find the zones of fast reconnection, we have employed an algorithm to identify the presence of current sheets in the turbulent regions (see \cite{zhdankin_etal_13,kadowaki_etal_18}) and computed statistically the magnetic reconnection rates in these locations (see \cite{kadowaki_etal_18}; for an application in shearing-box simulations).
	
\section{Identification of fast magnetic reconnection events}
\label{vrec_method}
	
\begin{figure}
  \begin{center}
    \includegraphics[scale=0.26]{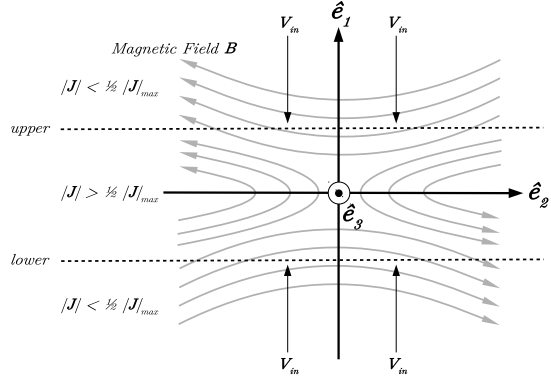}
	\caption{Scheme of the magnetic reconnection site in a new coordinate system. Obtained from \cite{kadowaki_etal_18}.}
	\label{fig:scheme}
  \end{center}
\end{figure}

We have applied the algorithm of Zhdankin et al. \cite{zhdankin_etal_13} and Kadowaki et al. \cite{kadowaki_etal_18} to identify magnetic reconnection sites in our simulations. Basically, we selected a sample of cells with a current density value higher than five times the average value taken in the whole system ($j_{max}>5\langle|\vec{\nabla}\times\vec{B}|\rangle$). Then, we constrained the sample and selected those cells where $j_{max}$ is the local maxima within a subarray of data, and located between magnetic field lines of opposite polarity. At each local maxima, we evaluated the eigenvalues and eigenvectors of the Hessian matrix of the current density to obtain a new coordinate system ($\hat{e}_1$,$\hat{e}_2$,$\hat{e}_3$) centered on the magnetic reconnection site (since it is not necessarily aligned with one of the axes of the Cartesian coordinate system; see Figure \ref{fig:scheme}). Finally, we evaluated the magnetic reconnection rate as in Kowal et al. \cite{kowal_etal_09}, where we have averaged the inflow velocity (to the reconnection region; $V_{in}$) divided by the Alfv\'{e}n speed ($V_{A}$) of each site identified by the algorithm (see \cite{kadowaki_etal_18}; for more details):
	
\begin{equation}
  \langle \frac{V_{in}}{V_{A}} \rangle = {1\over2} \left(\frac{V_{e_1}}{V_{A}}\vert_{lower} -\frac{V_{e_1}}{V_{A}}\vert_{upper} \right)~,
\end{equation}
where $V_{in}/V_{A}=V_{rec}$ is the magnetic reconnection rate, and $V_{e_1}$ is the projected velocity in the $\hat{e}_1$ direction at the lower and upper edges of the magnetic reconnection region\footnote{The edges of the magnetic reconnection region are defined by the cells with current density $|\vec{\nabla}\times\vec{B}|$ smaller than half of the maximum local value given by $j_{max}$ (see Figure \ref{fig:scheme})} (see Figure \ref{fig:scheme}).  

With this algorithm we can evaluate several features of the magnetic reconnection regions, such as the length and width of the reconnection structures, and the reconnected components of the magnetic field. Besides, the magnetic reconnection rate requires a careful analysis to remove false-positives sites identified by the algorithm (see \cite{kadowaki_etal_18} for a more detailed description of the overall method). In the next sections we will focus in the evaluation of $V_{rec}$ in GRMHD and SRMHD simulations.
	
\section{GRMHD accretion disk simulations}
	
We have used the ATHENA++ code \cite{white_etal_16} to perform global GRMHD simulations of a torus (thick disk, see \cite{fishbone_moncrief_76}) around a rotating black hole in a two-dimensional domain with $512$ cells in the radial direction and $512$ in the polar direction (in Kerr-Schild coordinates). We have assumed a black hole with mass $M=1$, spin $a/M = 0.95$, and adiabatic index $\Gamma=13/9$. The grid was set between $0.98$ times the outer horizon radius $M + \sqrt{M^2 - a^2}$ and $r=20$ in the radial direction, and between $\theta=0$ and $\theta=\pi$ in the polar direction. We have imposed outflow conditions in the radial boundaries and reflecting conditions in the polar boundaries. An LLF (local Lax-Friedrichs) Riemann solver was used.
	
\begin{figure}
  \begin{center}
    \includegraphics[scale=0.5]{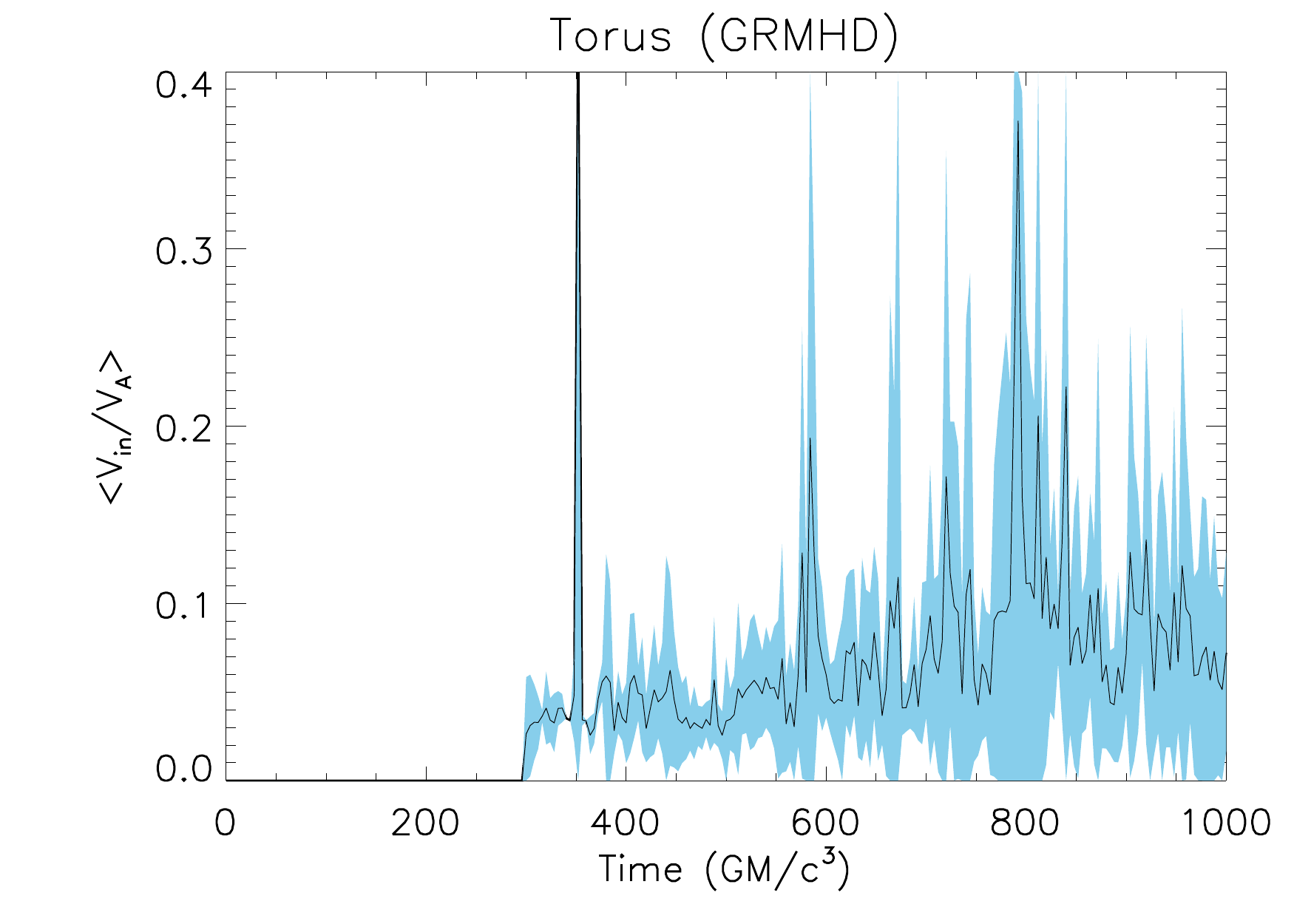}
	\includegraphics[scale=0.3]{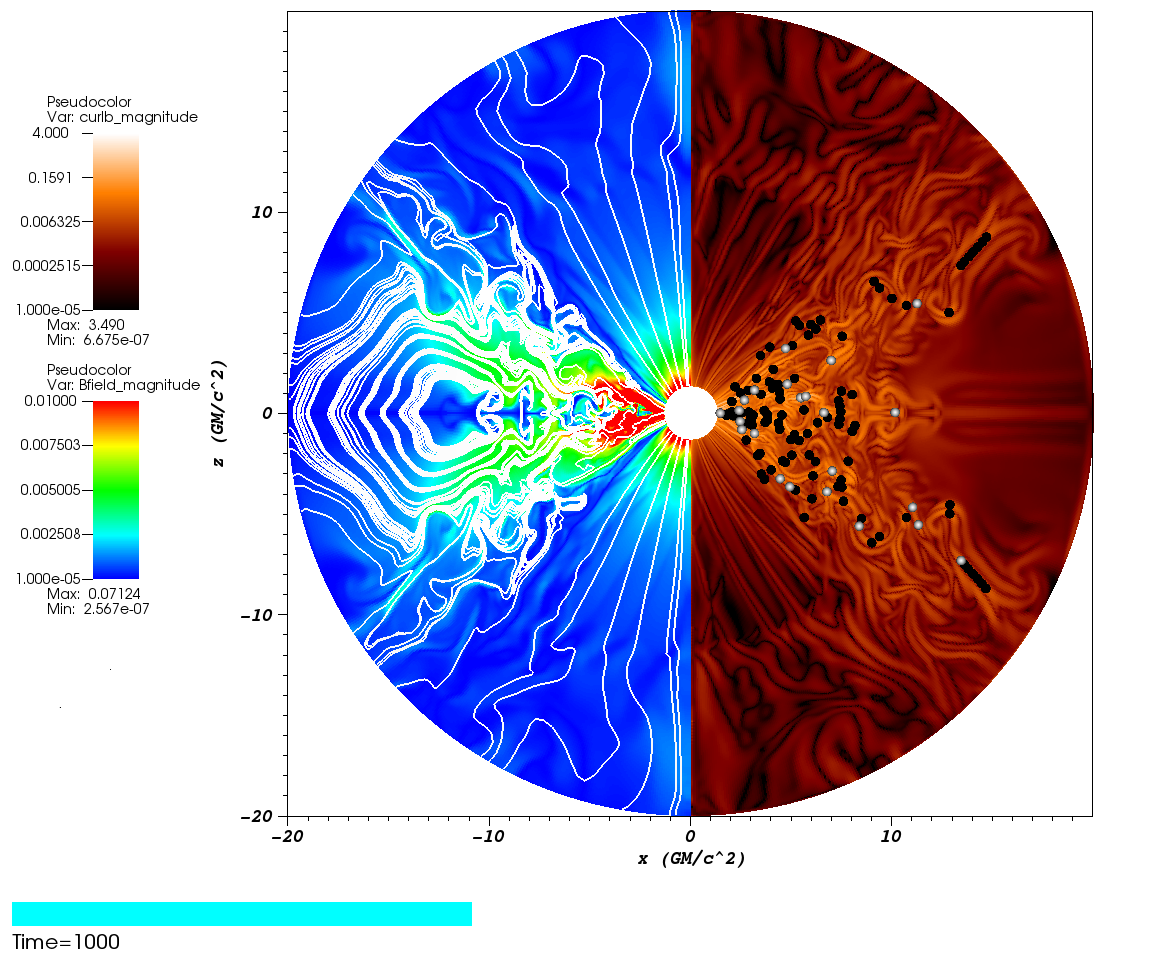}
	\caption{The top diagram shows the time evolution of the averaged magnetic reconnection (black line) measured by an observer in the coordinate frame. The blue shading corresponds to the standard deviation. The bottom diagram shows the system at $t = 1000$ (in units of $GM/c^3$). As time goes by the MRI sets in, allowing the accretion process and the formation of a turbulent environment. The black circles correspond to the local maxima identified by the algorithm and the white circles correspond to the confirmed magnetic reconnection sites (Kadowaki et al., in prep.).}
	\label{fig:vrec}
  \end{center}
\end{figure}

We have adapted the algorithm used in \cite{zhdankin_etal_13} and \cite{kadowaki_etal_18} to measure the magnetic reconnection rate \cite{kowal_etal_09} in a General Relativistic approach (see also \cite{ball_etal_18}). Figure \ref{fig:vrec} shows the magnetic reconnection rate measured by an observer at rest in the coordinate frame (top diagram) and the profiles of the magnetic field intensity and the current density (bottom diagram). This model corresponds to a torus with an initial weak poloidal magnetic field (represented by closed loops inside the torus) with the ratio of maximum gas pressure to maximum magnetic pressure equal $100$ (see \cite{white_etal_16}). The bottom diagram of Figure \ref{fig:vrec} shows, at $t = 1000$ (in units of $GM/c^3$), the formation of turbulent structures due to the MRI (that sets in at $t = 300$), allowing the accretion process and the development of magnetic reconnection sites (filled circle symbols). The black circles correspond to the local maxima current density identified by the algorithm, and the white circles correspond to the confirmed magnetic reconnection sites (see more details in \cite{kadowaki_etal_18}). The top diagram of Figure \ref{fig:vrec} shows the time evolution of the averaged values of $V_{rec}$ evaluated in the confirmed magnetic reconnection sites (white circles in the bottom diagram). We have obtained averaged values between $0.01$ and $0.7$ consistent with the predictions of the theory of turbulence-induced fast reconnection \cite{lazarian_vishiniac_99} (see also de Gouveia Dal Pino et al. 2019. in these Proceedings). 
	
\section{SRMHD jet simulations}
	
Recently, Singh, Mizuno and de Gouveia Dal Pino \cite{singh_etal_16} performed $3D$ relativistic MHD simulations of rotating Poynting-flux dominated jets with an initial helical geometry (suitable for relativistic jets near the launching region). Considering models with the ratio between the magnetic and the rest mass energy of the flow of the order of unit ($\sigma = 1$), and different density ratios between the jet and the environment, they induced precession perturbations that quickly developed current-driven kink (CDK) modes. This in turn induced turbulent and fast reconnection with rates of the order $0.05 V_A$\footnote{This value was obtained from an approximate version of equation 13 of \cite{kowal_etal_09}, whereas in the present work we have used the method described in Section \ref{vrec_method}}. We have also verified the efficiency of the magnetic reconnection in such systems since this process can play an important role in the particle acceleration along the jet (see de Gouveia Dal Pino et al. 2019, in these Proceedings). To this aim, we have extended the study above, using the special relativistic code Raishin \cite{mizuno_etal_06} in a $3D$ domain with $240$ cells in each direction. We have assumed an adiabatic index $\Gamma=5/3$ with an initial helical force-free configuration ($\vec{j}\times\vec{B}=0$) and the jet density profile higher than the environment (see \cite{singh_etal_16}). We have imposed outflow conditions in the $x$ and $y$ boundaries and periodic conditions in the vertical boundaries.
	
\begin{figure}
  \begin{center}
    \includegraphics[scale=0.2]{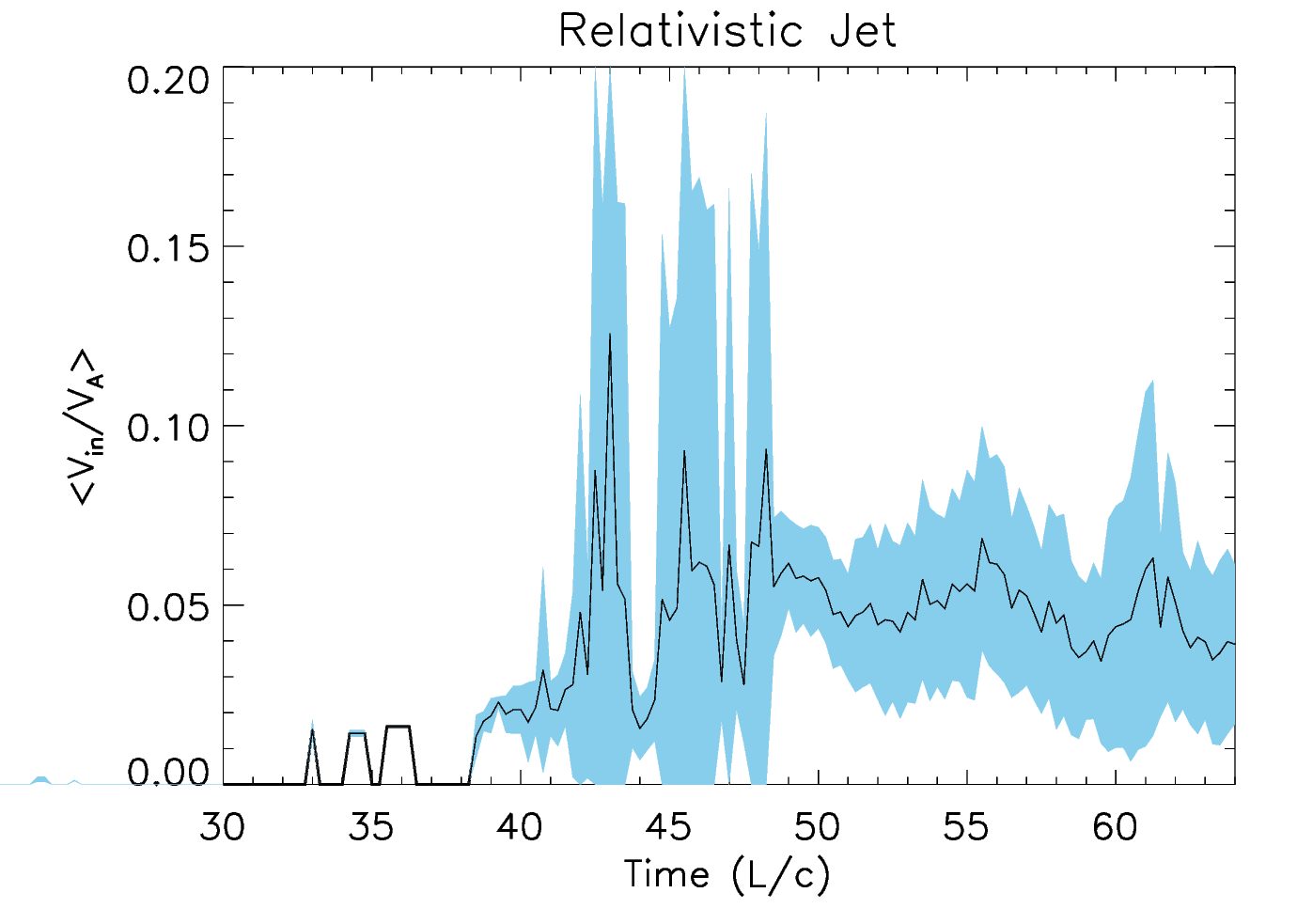}
	\includegraphics[scale=0.25]{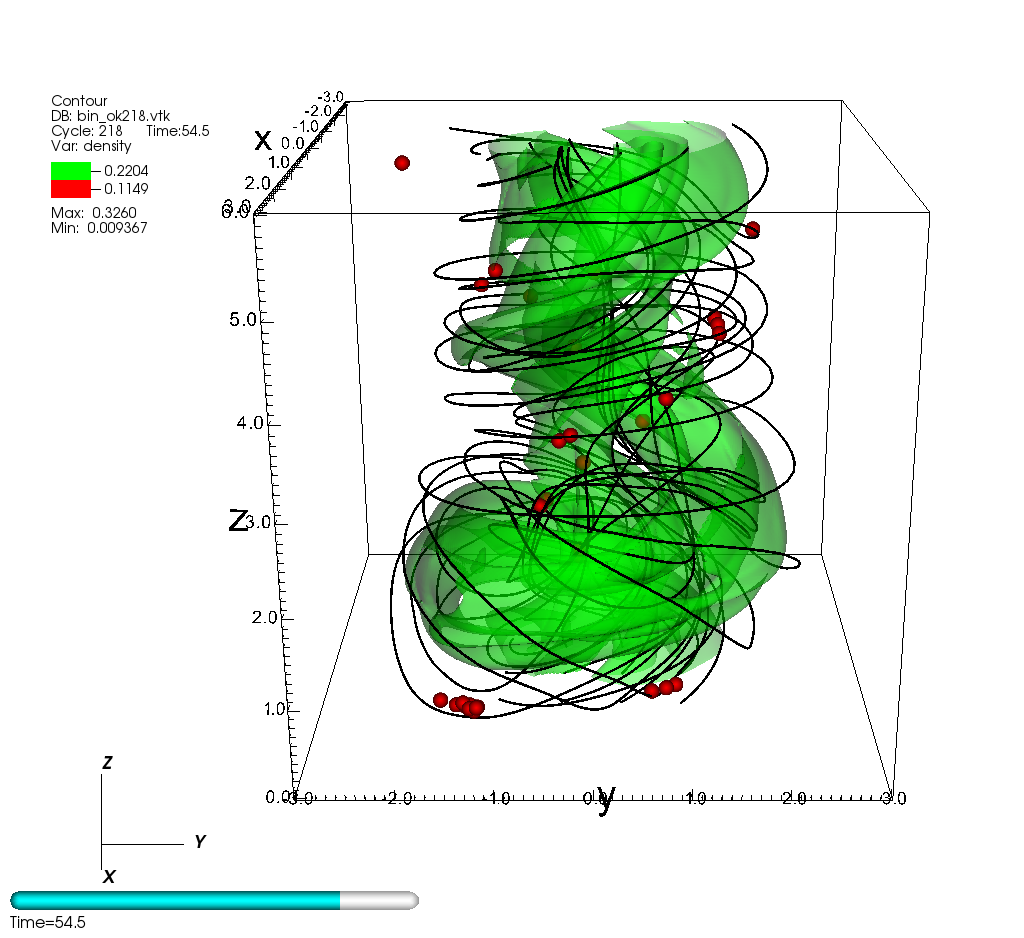}
	\caption{The top diagram shows the time evolution of the averaged magnetic reconnection rate (as in Figure \ref{fig:vrec}) along the relativistic helical jet. The bottom diagram shows the system at $t=54.5$ (in units of $L/c$). As time goes by the kink instability sets in, allowing the formation of magnetic reconnection sites (red points; more details in Kadowaki et al., in prep.).}
	\label{fig:reljet_vrec}
  \end{center}
\end{figure}

The bottom diagram of Figure \ref{fig:reljet_vrec} shows the density isosurface (green color) of the jet with the streamlines (black lines) of the magnetic field at $t=54.5 L/c$ (where $L$ is the longitudinal scale of the jet and $c$ the speed of light). The red points correspond to reconnection sites identified by the algorithm \cite{kadowaki_etal_18}. As in Figure \ref{fig:vrec}, the instability allows the development of magnetic reconnection events in several places along the jet. The top diagram of Figure \ref{fig:reljet_vrec} shows averaged values for the reconnection rate between $0.01$ and $0.15$ and indicates that multiple fast reconnection events take place due to the CDK instability (that sets in at $t = 38 L/c$).
	
\begin{figure}
  \begin{center}
    \includegraphics[scale=0.4 ]{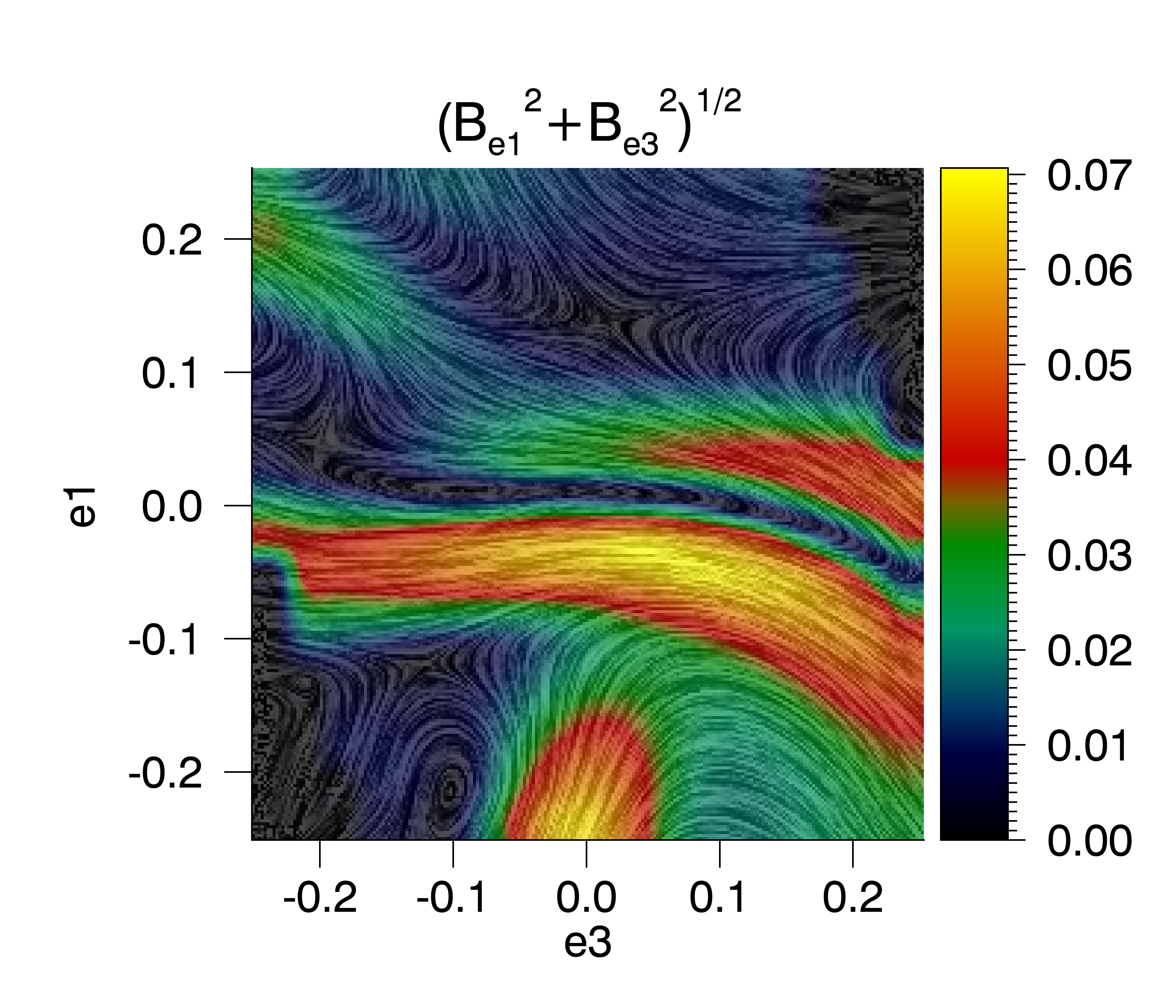}
    \includegraphics[scale=0.4 ]{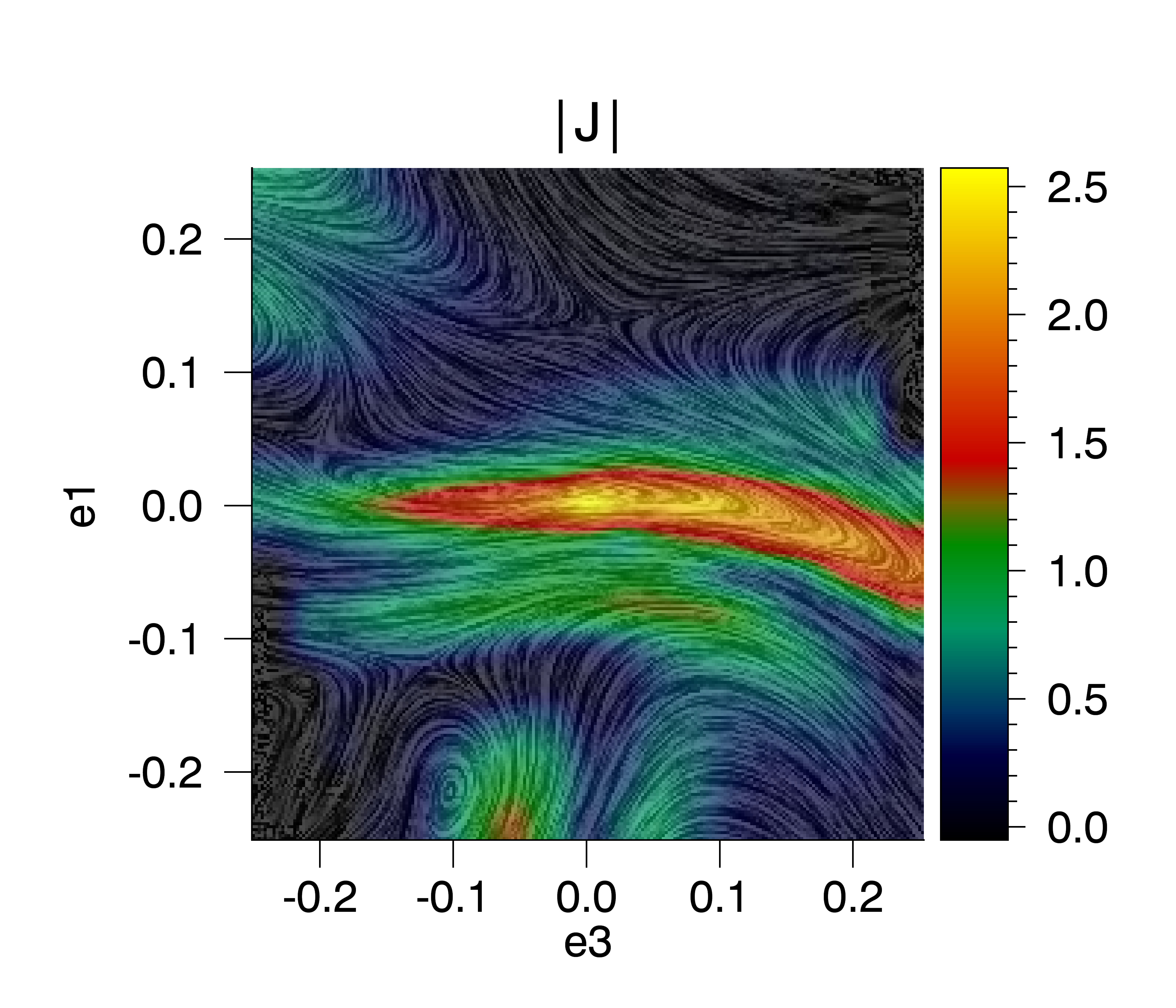}
	\includegraphics[scale=0.4 ]{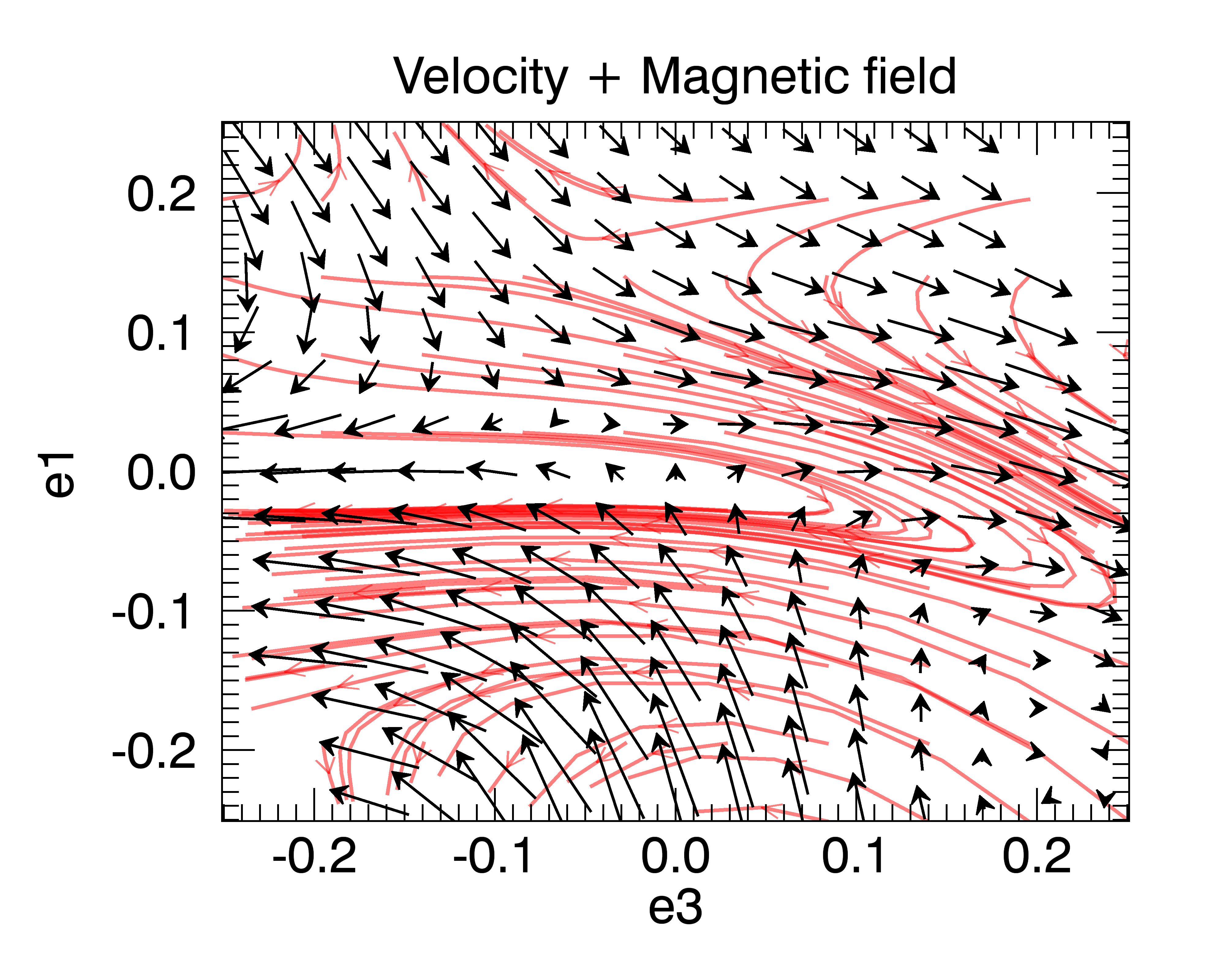}
	\caption{Diagrams showing a magnetic reconnection region of the domain produced by a line integral convolution method combined with the maps of the magnetic field (top left diagram) and total current density (top right diagram). The bottom diagram shows the streamlines of the magnetic field (red lines) and the velocity field (black arrows). The diagrams have been obtained by interpolating the data in the surroundings of the reconnection region identified by the algorithm (Kadowaki et al., in prep.).}
  \end{center}
  \label{fig:reljet_lic}
\end{figure}

Figure \ref{fig:reljet_lic} shows an example of one of the reconnection sites identified by the algorithm at $t=54.5 L/c$. The $2D$ maps of the magnetic field intensity (top left diagram) and current density (top right diagram) have been produced using the line integral convolution (LIC) method \cite{cabral_leedom_93}. The bottom diagram shows the streamlines of the magnetic field (red lines) and the velocity field (black arrows). The diagrams were obtained by a cubic interpolation of the data in the surroundings of the reconnection site (originally the cubic data had 21 cells in each direction and was increased ten times for visualization purposes) and correspond to an arbitrary slice in the $\hat{e}_{1}-\hat{e}_{3}$ plane (in the local coordinate system centered on the reconnection site). The magnetic and velocity fields were evaluated using the $\hat{e}_{1}$ and $\hat{e}_{3}$ components. As we expected, it is clear the anti-correlation between the magnetic field intensity and the current density. The bottom diagram shows the magnetic field lines pushed by the velocity flow performing a reconnection event. We also see converging velocity field of the fluid arising from the reconnection site (as in Figure \ref{fig:scheme}).  
	
\section{Summary}
	
In this work, we have employed an algorithm to identify the presence of fast magnetic reconnection events, presenting preliminary results for the surroundings a torus around a black hole (built in GRMHD simulations) and along helical Poynting-flux jets (built in SRMHD simulations). We have computed the magnetic reconnection rates in these locations and detected the presence of fast reconnection events, obtaining average values of the order of $0.01$ and $0.7$ (in the torus) and $0.01$ and $0.15$ (in the jet)\footnote{Currently, we are concluding convergence studies to verify the dependence of $V_{rec}$ with resolution. For both cases (accretion torus and the jet), we have models with twice the resolution of the models presented here that confirm the results.}, as predicted by the theory of turbulence-induced fast reconnection \cite{lazarian_vishiniac_99}. This result strengthens the scenario where turbulent fast magnetic reconnection can take place in astrophysical environments such as the core region of BHBs and AGNs (see \cite{dgdp_lazarian_05,dgdp_etal_10,kadowaki_etal_15,singh_etal_15,rodriguezramires_etal_18}) and relativistic jets, where the magnetic energy released by these events can accelerate relativistic particles and produce non-thermal emissions observed in these sources (see de Gouveia Dal Pino et al. 2019, in these proceedings).
	
\acknowledgments
The numerical simulations in this work have been performed in the Blue Gene/Q supercomputer supported by the Center for Research Computing (Rice University) and Superintend\^{e}ncia de Tecnologia da Informa\c{c}\~{a}o da Universidade de S\~{a}o Paulo (USP). This work has also made use of the computing facilities of the Laboratory of Astroinformatics (IAG/USP, NAT/Unicsul), whose purchase was made possible by the Brazilian agency FAPESP (grant 2009/54006-4) and the INCT-A; and the cluster of the group of Plasmas and High-Energy Astrophysics (GAPAE), acquired also by the Brazilian agency FAPESP (grant 2013/10559-5). LHSK acknowledges support from the Brazilian agency FAPESP (postdoctoral grant 2016/12320-8), and EMGDP from FAPESP (grant 2013/10559-5) and CNPq (grant 308643/2017-8).

\end{document}